\documentstyle[12pt]{article}
\begin{document}
\input FEYNMAN
\textheight 550pt
\textwidth 300pt
\addtolength{\topmargin}{-0.8in}
\addtolength{\textheight}{1in}
\hsize=16.5truecm
\hoffset=-.5in
\baselineskip=7mm

\thispagestyle{empty}
\begin{flushright} \
KHTP-94-02 \\
SNUTP 94-27 \\
hep-th/9412173 \\
\end{flushright}

\begin{center}
 {\large\bf  An Effective Action for Quasi-elastic Scatterings in QCD }\\[.3in]
B.K. Chung, Soonkeon Nam,
\footnote{E-mail address: nam@nms.kyunghee.ac.kr}
Q-Han Park,
\footnote{E-mail address: qpark@nms.kyunghee.ac.kr}
and H.J. Shin
\footnote{E-mail address: hjshin@nms.kyunghee.ac.kr}
\\[.5in]
{\it  Department of Physics\break
	   and\break
       Research Institute for Basic Sciences\break
	  Kyung Hee University\break
	  Seoul, 130-701, Korea}
\\[.3in]
{\bf ABSTRACT }
\end{center}
\vskip .2in
A new effective action for the high energy quark-quark scatterings is
obtained by applying a scaling approximation to the QCD action.
The propagators are shown to factorize into the transverse and the
longitudinal parts
so that the scattering amplitudes are given in terms of the products of two
dimensional $S$-matrices.
we show that our action provides a natural effective field theory for  the
Lipatov's theory of quark scatterings with quasi-elastic unitarity.
The amplitude with quasi-elastic unitarity obtained from this action
shows `Regge' behavior and is eikonalized.
\newpage
\def\a{\alpha}
\def\b{\beta}
\def\l{\lambda}
\def\beq{\begin{equation}}
\def\eeq{\end{equation}}
\def\bea{\begin{eqnarray}}
\def\eea{\end{eqnarray}}
\renewcommand{\arraystretch}{1.5}
\def\ba{\begin{array}}
\def\ea{\end{array}}
\def\bce{\begin{center}}
\def\ece{\end{center}}
\def\nn{\noindent}
\def\pbx{\partial_x}
\def\pnot{\not{p}}
\def\knot{\not{k}}

\def\ptl{\partial}
\def\al{\alpha}
\def\be{\beta}
\def\ga{\gamma}
\def\Ga{\Gamma}
\def\de{\delta} \def\De{\Delta}
\def\ep{\epsilon}
\def\e{\epsilon}
\def\vep{\varepsilon}
\def\ze{\zeta}
\def\et{\eta}
\def\th{\theta} \def\Th{\Theta}
\def\vth{\vartheta}
\def\io{\iota}
\def\ka{\kappa}
\def\la{\lambda}
\def\La{\Lambda}
\def\rh{\rho}
\def\si{\sigma} \def\Si{\Sigma}
\def\ta{\tau}
\def\up{\upsilon}
\def\Up{\Upsilon}
\def\ph{\phi}
\def\Ph{\Phi}
\def\vph{\varphi}
\def\ch{\chi}
\def\ps{\psi}
\def\Ps{\Psi}
\def\om{\omega}
\def\Om{\Omega}
\def\para{\parallel}

\def\lbr{\left(}
\def\rbr{\right)}
\def\ben{\begin{eqnarray}}
\def\een{\end{eqnarray}}

\def\vol#1{{\bf #1}}
\def\MPL#1{Mod. Phys. Lett. \vol{#1} }
\def\nupha#1{Nucl. Phys. \vol{#1} }
\def\phlta#1{Phys. Lett. \vol{#1} }
\def\phyrv#1{Phys. Rev. \vol{#1} }
\def\PRL#1{Phys. Rev. Lett \vol{#1} }
\def\prs#1{Proc. Roc. Soc. \vol{#1} }
\def\PTP#1{Prog. Theo. Phys. \vol{#1} }
\def\JETP#1{Sov. Phys. JETP \vol{#1} }
\def\SJNP#1{Sov. J. Nucl. Phys. \vol{#1} }
\def\TMP#1{Theor. Math. Phys. \vol{#1} }
\def\ANNPHY#1{Annals of Phys. \vol{#1} }


It  is believed that  perturbative amplitudes of high-energy scatterings in
QCD for the large value of the Mandelstam variable $s$ and
fixed value of $t$ could be expressed in terms of two dimensional amplitudes.
This raised the possibility of an effective description of high
energy scatterings in QCD in the context of two-dimensional field theories.
Recently, there have been some efforts in this direction, notably by
Lipatov\cite{Lipatov} and Verlinde's\cite{Verlinde}.
In particular,  the two-dimensional nature appears
in an approximation scheme introduced by Lipatov to construct an
$S$-matrix which preserves  $s$ and $u$ channel unitarity.
In QCD,  the leading log approximation\cite{LLA}(LLA) gives rise to the
Reggeized gluon propagator. Consequently, the total cross section rises as
$\si_{\rm tot}\sim s^{\De}$ which  violates the Froissart bound of
$\si_{\rm tot}< c\ln^{2} s$.
This violation results in the breakdown of unitarity in LLA.
To remedy the problem, Lipatov et al. suggested a modification of
LLA - known as  the $i\pi$-approximation\cite{Lipatov}.
Alternatively, Lipatov suggested a field theoretic approach in calculating
the scattering amplitudes with quasi-elastic unitarity.
This approximation, named as ``the quasi-elastic unitarity(QEU)
approximation"\cite{QuasiEla}, takes into account
only soft gluons with small transverse momenta in the
intermediate states of the $s$ and $u$ channels in calculating the $S$-matrix.
In the QEU approximation for the quark-quark scattering,
the scattering amplitude for the box diagram has been  calculated
by assuming  the quark propagators of the following form:
\beq
   {\cal D}_{-}=\frac{1}{k_{-}+i\ep/k_{+}}, \quad
   {\cal D}_{+}=\frac{1}{k_{+}+i\ep/k_{-}}, \quad k_{\pm}=k_{0}\pm k_{3},
\eeq
for incident quarks with momenta $k_{\pm }$.
The gluon interactions are described by the
four particle vertices $G$,
\beq
  G(M\rh)=\frac{g^{2}}{(2\pi)^{3}}\int_{|q|^{2}\gg M^{2}}
  \frac{d^{2}q}{|q|^{2}}{\rm e}^{-iq\rh} = \frac{g^{2}}{8\pi^{2}}\ln
  \frac{1}{\rh^{2} M^{2}}.
\eeq
where $M$ is IR cutoff and $\rh$ is the impact parameter.
Higher order scattering amplitudes are obtained by  the
unitarity condition and dispersion relations up to the subtraction terms.
These subtraction terms have been determined  through
the evolution equation and    the explicit evaluation of the box diagrams
using Eqs.(1) and (2).

The purpose of this letter is to explain the QEU approximation in the context
of
full QCD action and to show  the `Regge' behavior and  eikonalization of
the amplitudes in the  QEU approximation.  In particular, we propose an
effective action of two dimensional nature, obtained from the Yang-Mills
action by applying the scaling argument of Verlinde's\cite{Verlinde}.
In this theory,  we show that the  fermionic propagator is purely
longitudinal while the bosonic propagator
factorizes into the transverse and the longitudinal parts.
The transverse part after integration
becomes an effective four particle coupling constant $G$ thereby making the
theory effectively two dimensional.  This provides a field theoretic
explanation for  Lipatov's QEU approximation.
We also compute the higher order diagrams perturbatively in  the coupling
constant $G$. Remarkably, these perturbative terms may be summed over for
each irreducible isospin factors which is effectively equivalent
to replacing the coupling  constant $G$ by the `Reggeized'  coupling
$\tilde{G} = G/(1 - {1 \over 2} c_{v} G\ln s)$ where $c_{v}$ is the quadratic
Casimir in the adjoint
representation of the color group.
In the restriction to the abelian case, the scattering
amplitude manifests the eikonalizing behavior.

In order to describe the high energy forward scatterings, we consider the
the scaling argument of Verlinde's which is given by rescaling of the
longitudinal coordinates $x_\al$ and transverse coordinates
$x_i$\cite{Verlinde};
\beq
x^{\al } \rightarrow \l x^{\al} \ , \ x^{i} \rightarrow x^{i},  \ \ \
{\rm with}\ \ \  A_{\al } \rightarrow \l^{-1}A_{\al} \ ,
\ A_{i} \rightarrow A_{i}.
\eeq
This separates the Yang-Mills action into the strongly and the weakly
coupled parts;
\beq
S_{\rm YM}={1\over 2}\int \mbox{tr} (E^{\al\b}F_{\al\b} +F_{\al i}F^{\al i})+
{\l^{2} \over 4}\int \mbox{tr} (E_{\al\b}E^{\al\b } + F_{ij}F^{ij}),
\eeq
where the auxiliary field $E_{\al\b}$ is introduced to avoid the
singular behavior of $ \la^{-2 }F_{\al \b}F^{\al \b} $ in the action.
The high energy limit is obtained by $\l \rightarrow 0 $, where
the Yang-Mills action  becomes
\beq
S_{1} = {1\over 2}\int \mbox{tr} (E^{\al\b}F_{\al\b} +F_{\al i}F^{\al i}),
\eeq
whereas the truncated quark action takes the form
\beq
S_{2} = \int \bar{\psi}\gamma^{\al}(\partial_{\al }+ gA_{\al })\psi.
\eeq
The Lagrange multiplier $E_{\al\b }$, when eliminated, requires
$F_{\al \b } = 0$ which we solve for   $gA_{\al}=\partial_{\al}UU^{-1}$
so that Eq.(5) becomes
\beq
S_{1}[U,A_{i}]=\frac{1}{2g^{2}}\int {\rm tr} [\partial_{\al}(U^{-1}
(\partial_{i}+ gA_{i})U)]^{2}.
\eeq
Instead of using $A_{i}$ and $U$, we make field redefinitions;
$a_{i} \equiv U^{-1}A_{i}U $ and $U\equiv \exp(g\ph)$. We fix the gauge by
$\partial_{i}a_{i} = 0$ which we solve in terms of a scalar field $\alpha $
such that $a_{i} = \e _{ij}\partial_{j}\al$.
Notice that  our field redefinition respects the scaling approximation,
i.e. $a_{i} $ is invariant under the scaling Eq.(3), which is necessary in
order not to mix strongly and weakly coupled parts of Eq.(4). Field
redefinition, for example, $ a_{i} = U^{-1}(\partial_{i} + gA_{i})U$ which
makes the transverse part  $A_{i}$ completely decouple from the action,
breaks the scaling approximation and thus is not allowed. The Jacobian of
changing variables from $A_{i}$ to $\al $ is simply one.  The virtue of
using $\al $ is that it makes $\ph $ and $\al $ decouple in the lowest order
of the action,
\bea
S_{1}&=&\int {\rm tr}\{\frac{1}{4}\ptl_{+}\ptl_{i}\ph\ptl_{-}\ptl_{i}\ph+
	  \frac{1}{4}\ptl_{+}\ptl_{i}\al\ptl_{-}\ptl_{i}\al
	  + \frac{g}{4} \ptl_{+}\ptl_{i}\ph\ptl_{-}[\ptl_{i}\ph,\ph]  -
	  \frac{g}{4}\ptl_{+}\al\ptl_{-}[\ptl_{i}\ph,\ep_{ij}\ptl_{j}\ph]
	  \nonumber \\
& & + \frac{g^2}{12}\ptl_{+}\ptl_{i}\ph\ptl_{-} [[ \ptl_{i}\ph , \ph ], \ph ]
+\frac{g^2}{16}\ptl_{+} [ \ptl_{i} \ph , \ph ] \ptl_{-} [\ptl_{i} \ph , \ph ]
-\frac{g^2}{12}\ptl_{+}\al \e_{ij}\ptl_{j}\ptl_{-}[[\ptl_{i} \ph , \ph ] ,
\ph ] \nonumber \\
&& + {\cal{O}}(g^3) \}  + \{  + \leftrightarrow - \},
\eea
and
\beq
S_{2} = \int \bar{\psi}\gamma^{\al}(\partial_{\al }+ g \ptl_{\al }\ph  +
{\cal{O}}(g^2) )  \psi \ .
\eeq
Also, the propagators of $\ph $ and $\al $ factorize into the transverse and
the longitudinal parts;
\beq
{1 \over (p_{\perp }^{2} + i \e )( p_{\para }^{2} + i\e )}.
\eeq
The factorization of propagators is an essential feature in the
Lipatov's QEU approximation.
We demonstrate this through the calculation of the box diagram for
the $u$ and $s$-channel in the $S$-matrix.
The quark-quark scattering amplitude $T$ is given as follows,
with the kinematic factors $x,y$ are given by $x = \ln{-\La/ s}$,
$y = \ln { \La/ s}$, and
\vskip -0.2in
\hskip 0.5in
\begin{picture}(10000,10000)
\THICKLINES\drawline\fermion[\E\REG](0,0)[3000]
\THICKLINES\drawarrow[\W\ATBASE](\pmidx,\pmidy)
\THICKLINES\drawline\fermion[\E\REG](0,4000)[3000]
\THICKLINES\drawarrow[\E\ATBASE](\pmidx,\pmidy)
\put(-3000,1900){$T=$}

\put(3300,2000){$~~+~~2 \pi i G$}
\THICKLINES\drawline\fermion[\E\REG](9000,0)[3000]
\thinlines\drawline\fermion[\N\REG](\pmidx,\pmidy)[4000]
\THICKLINES\drawline\fermion[\E\REG](9000,4000)[3000]
\put(12500,2000){~$+~~2 \pi i G^2~\Biggl($}
\put(24000,2000){$x~~+$}
\put(33000,2000){$(-y)\Biggr)$}

\THICKLINES\drawline\fermion[\E\REG](19500,0)[3500]
\thinlines\drawline\fermion[\N\REG](20500,0)[4000]
\thinlines\drawline\fermion[\N\REG](22000,0)[4000]
\THICKLINES\drawline\fermion[\E\REG](19500,4000)[3500]

\THICKLINES\drawline\fermion[\E\REG](27500,0)[5000]
\thinlines\drawline\fermion[\NE\REG](28000,0)[5500]
\thinlines\drawline\fermion[\NW\REG](32000,0)[5500]
\THICKLINES\drawline\fermion[\E\REG](27500,4000)[5000]
\put(36500,2000){$+~\cdots$}

\put(15000,-4000){Fig. 1}
\end{picture}
\vskip 1in
 The box diagram in  the $s$-channel  in Fig.1 gives rise to
$ \int d^2 k_{\para } d^2 k_{\perp } A_{\mbox{fermi}}A_{\mbox{gluon}}
{\cal G}_{2} \equiv f_{2}{\cal G}_{2} $ where ${\cal G}_{2}$
is the group factors
${\cal G}_{2}\equiv (\sum_a T_{a}^{L} T_{a}^{R} )^2 $
and
\beq
A_{\mbox{fermi}} =  { \bar{u}(p_{1})\gamma^{+}(\pnot_{1} - \knot )\gamma^{+}
u(p_{1}) \over [(p_{1}-k)^2 + i\e ] } \times
 { \bar{u}(p_{2})\gamma^{-}(-\pnot_{2} - \knot )\gamma^{-} u(p_{2})
 \over [(-p_{2}-k)^2 + i\e ] }
\eeq
and
\beq
A_{\mbox{gluon}} = { k_{\para }^2 ( q_{\para }- k_{\para })^2 \over
k_{\perp}^2 k_{\para}^2 (q_{\perp}-k_{\perp })^2 (q_{\para}-k_{\para })^2} \ .
\eeq
We have chosen the external momenta of quarks to be $p_{1}= (w, 0 ,   0, -w),
p_{2}=(w,0,0,w)$ or $ p_{1+} = p_{2-}= 0$ so that $ \bar{u}(p_{1})
\gamma^{-}u(p_{1})   = p_{1+} = 0 \ , \
\bar{u}(p_{2}) \gamma^{+}u(p_{2}) =  p_{2-} = 0$.
The numerator in $A_{\mbox{gluon}}$ comes from the interacting vertex
in Eq.(9). Therefore $f_{2}$ reduces to
\beq
f_{2} =  \int {dk_{+}dk_{-}  \over [(p_{1} -k)_{-} +
{ i\e \over (p_{1} - k)_{+}}][(-p_{2} -k)_{+} +
{ i\e \over (-p_{2} - k)_{-}}] } \int {d^2k_{\perp}
\over  k_{\perp}^2  (q_{\perp }-k_{\perp })^2  } (p_{1} + p_{2})^2 .
\eeq
Note that the $k_{+}(k_{-})$-integral in Eq.(13)  agrees with the
Lipatov's QEU approximation for  a diagrammatic representation of the
scattering matrix. Higher order Feynman
diagrams from the action Eqs.(8) and (9) may be computed similarly.
Here, we restrict ourselves  only to subsets of
higher order digrams which appear in the QEU approximation.
These higher order diagrams can be computed recursively from the box
diagram by using the dispersion and the unitarity conditions without
explicit evaluation of each Feynman
diagrams\cite{QuasiEla}. The result for the scattering amplitude is
\beq
T = 1 + 2 \pi i \sum_{n=1}^\infty G^n ( P_{n-1} (x) + Q_{n-1} (y) ),
\eeq
and
\begin{eqnarray}
P_0 &=& Q_0 = {1 \over 2} [1] \nonumber \\
P_1(x) &=& [1 2] x, \ \ Q_1(y) = -  P_1^T (y) = [2 1] (-y)
\nonumber \\
P_2(x) &=& [1 2 3] x^2 + ([1 3 2] + [2 1 3]) (-{1 \over 2} x^2 - {1 \over 6}
\pi^2), \ \ Q_2(y) =  {(P_2^T (y))}
\nonumber \\
P_3(y) &=& [1 2 3 4] x^3 + ([2 1 3 4] + [1 2 4 3]) (-{1 \over 2} x^3 -
{1 \over 6} \pi^2 x ) + [1 3 2 4] (-{1 \over 3} x^3 + {1 \over 3} \pi^2
x)
\nonumber \\
&+& ([2 3 1 4] + [3 1 2 4] + [1 3 4 2] + [1 4 2 3]) ( -{1 \over 6} x^3
- {1 \over 3} \pi^2 x ) \nonumber \\
& +&  ([3 2 1 4] + [2 1 4 3] + [1 4 3 2])({
1 \over 3} x^3 + {1 \over 3} \pi^2 x) \ , \
Q_3 (y) = -  P_3^T (y) \ .
\end{eqnarray}
The permutation  $[n_1 , n_2 , \cdots ] $ of a set of positive
consecutive integers $(1,2, \cdots )$ represents an isospin factor of
permutation type
\beq
[n_1 , n_2 , \cdots ] \equiv \sum_{a_1, a_2,\cdots = 1}^{\dim G}
T^{L}_{a_1} T^{L}_{a_{2}}\cdots T^{R}_{a_{n_1}} T^{R}_{a_{n_2}} \cdots ,
\eeq
for example, $[213] = \sum_{a,b,c} T^{L}_{a} T^{L}_{b}T^{L}_{c}
  T^{R}_{b} T^{R}_{a}T^{R}_{c}$.
In the above, $T^R_a$  and $T^L_a$ are the nonabelian charges of the left
and right moving quarks respectively.
$P^{T}(y)$ is the same as $P(y)$ except the isospin factors
$[n_1 , n_2 , \cdots , n_i] $ are replaced by the reversly ordered ones
$[n_i  , \cdots  , n_2 , n_1 ]$.
However, the isospin factors in Eq.(16) are not unique; they can be
rearranged in terms of other isospin factors. For example
\bea
[21] &=& \sum_{a,b}T_{a}^{L}T_{b}^{L}T_{b}^{R}T_{a}^{R} =
\sum_{a,b}T_{a}^{L}T_{b}^{L}(T_{a}^{R}
T_{b}^{R}+ if_{bac}T_{c}^{R}) \nonumber \\
&=&  \sum_{a,b}T_{a}^{L}T_{b}^{L}T_{a}^{R}T_{b}^{R} +
{c_{v} \over 2}\sum_{a}T_{a}^{L}T_{a}^{R}
= [12] +  {c_{v} \over 2}[1].
\eea
Similarly, higher order isospin factors can also be rearranged, e.g. the
isospin factor  $[3412]$ can be depicted pictorially by the following
diagram:

\hskip 1in
\begin{picture}(10000,10000)
\THICKLINES\drawline\fermion[\E\REG](-10000,0)[7000]
\thinlines\drawline\fermion[\NE\REG](-9500,0)[5500]
\thinlines\drawline\fermion[\NE\REG](-8000,0)[5500]
\thinlines\drawline\fermion[\NW\REG](-5000,0)[5500]
\thinlines\drawline\fermion[\NW\REG](-3500,0)[5500]
\THICKLINES\drawline\fermion[\E\REG](-10000,4000)[7000]
\put(-8000,5000){$[3412]$}

\THICKLINES\drawline\fermion[\E\REG](1000,0)[5000]
\thinlines\drawline\fermion[\N\REG](2000,0)[4000]
\thinlines\drawline\fermion[\N\REG](3000,0)[4000]
\thinlines\drawline\fermion[\N\REG](4000,0)[4000]
\thinlines\drawline\fermion[\N\REG](5000,0)[4000]
\put(-1000,2000){$=$}
\THICKLINES\drawline\fermion[\E\REG](1000,4000)[5000]
\put(2000,5000){$[1234]$}

\THICKLINES\drawline\fermion[\E\REG](11500,0)[4000]
\thinlines\drawline\fermion[\N\REG](12500,0)[4000]
\thinlines\drawline\fermion[\N\REG](13500,0)[4000]
\thinlines\drawline\fermion[\N\REG](14500,0)[4000]
\put(7000,2000){$~+~~2c_v$}
\THICKLINES\drawline\fermion[\E\REG](11500,4000)[4000]
\put(12500,5000){$[123]$}

\THICKLINES\drawline\fermion[\E\REG](22000,0)[4000]
\thinlines\drawline\fermion[\N\REG](23000,0)[4000]
\thinlines\drawline\fermion[\N\REG](25000,0)[4000]
\put(17000,2000){$~+~~\frac{1}{2}c_v^2$}
\THICKLINES\drawline\fermion[\E\REG](22000,4000)[4000]
\put(23000,5000){$[12]$}

\THICKLINES\drawline\fermion[\E\REG](32000,0)[4000]
\thinlines\drawline\fermion[\N\REG](33000,0)[4000]
\thinlines\drawline\fermion[\N\REG](35000,0)[4000]
\thinlines\drawline\fermion[\E\REG](33000,2000)[2000]
\put(26000,2000){$~~~-~~\frac{1}{2}c_v$}
\THICKLINES\drawline\fermion[\E\REG](32000,4000)[4000]
\put(33000,2000){\circle*{300}}
\put(35000,2000){\circle*{300}}
\put(33000,5000){$[H]$}

\THICKLINES\drawline\fermion[\E\REG](1500,-9000)[5000]
\thinlines\drawline\fermion[\N\REG](2500,-9000)[4000]
\thinlines\drawline\fermion[\E\REG](2500,-7000)[1500]
\thinlines\drawline\fermion[\N\REG](4000,-9000)[4000]
\thinlines\drawline\fermion[\N\REG](5500,-9000)[4000]
\THICKLINES\drawline\fermion[\E\REG](1500,-5000)[5000]
\put(2500,-7000){\circle*{300}}
\put(4000,-7000){\circle*{300}}
\put(-500,-7000){$~-~~$}
\put(2500,-4000){$[H1]$}

\THICKLINES\drawline\fermion[\E\REG](10000,-9000)[5000]
\thinlines\drawline\fermion[\N\REG](11000,-9000)[4000]
\thinlines\drawline\fermion[\E\REG](12500,-7000)[1500]
\thinlines\drawline\fermion[\N\REG](12500,-9000)[4000]
\thinlines\drawline\fermion[\N\REG](14000,-9000)[4000]
\THICKLINES\drawline\fermion[\E\REG](10000,-5000)[5000]
\put(12500,-7000){\circle*{300}}
\put(14000,-7000){\circle*{300}}
\put(8000,-7000){$~-~~$}
\put(11000,-4000){$[H2]$}

\THICKLINES\drawline\fermion[\E\REG](18500,-9000)[5000]
\thinlines\drawline\fermion[\N\REG](19500,-9000)[4000]
\thinlines\drawline\fermion[\E\REG](19500,-7000)[3000]
\thinlines\drawline\fermion[\N\REG](21000,-9000)[4000]
\thinlines\drawline\fermion[\N\REG](22500,-9000)[4000]
\THICKLINES\drawline\fermion[\E\REG](18500,-5000)[5000]
\put(19500,-7000){\circle*{300}}
\put(22500,-7000){\circle*{300}}
\put(16100,-7000){$~-~~$}
\put(19500,-4000){$[H3]$}

\THICKLINES\drawline\fermion[\E\REG](27000,-9000)[5500]
\thinlines\drawline\fermion[\N\REG](28500,-9000)[4000]
\thinlines\drawline\fermion[\E\REG](28500,-7000)[2500]
\thinlines\drawline\fermion[\N\REG](31000,-9000)[4000]
\thinlines\drawline\fermion[\NW\REG](31700,-9000)[5500]
\THICKLINES\drawline\fermion[\E\REG](27000,-5000)[5500]
\put(31000,-7000){\circle*{300}}
\put(28500,-7000){\circle*{300}}
\put(25000,-7000){$-~~$}
\put(28500,-4000){$[H4]$}

\put(14000,-14000){Fig 2}
\end{picture}
\vskip 2in
Since each vertex carries $if_{abc}$, the isospin factors of
non-permutation type,
e.g. the $[H]$-diagram is defined as follows:
\vskip 1in
\hskip 1in
\begin{picture}(0,0)
\THICKLINES\drawline\fermion[\E\REG](0,0)[4000]
\thinlines\drawline\fermion[\N\REG](1000,0)[4000]
\thinlines\drawline\fermion[\N\REG](3000,0)[4000]
\thinlines\drawline\fermion[\E\REG](1000,2000)[2000]
\put(5000,2000){$~~~= ~~ \sum_{abcde}
(if_{abd})(if_{aec})T^{L}_{b}T^{L}_{c}T^{R}_{d}T^{R}_{e}$}
\THICKLINES\drawline\fermion[\E\REG](0,4000)[4000]
\put(1000,2000){\circle*{300}}
\put(3000,2000){\circle*{300}}
\end{picture}
\vskip 2in
Other isospin factors of $H$-type appearing in Fig.2 can be written down
easily by attaching extra $T_a$ factors to the $H$-diagram\cite{ChengWu}.
By rearranging diagrams in Eq.(15), we have an equivalent expression for the
scattering amplitude in terms of irreducible isospin factors,
\bea
T &=& 1 + 2\pi i[1] (G-\frac{c_v}{2} G^2 y+\frac{c_v^2}{4}
G^3 y^2-\frac{c_v^3}{8} G^4 y^3 ) + 2\pi i [12] (-i\pi) (G^2-c_v G^3 y+
\frac{3}{4} c_v^2 G^4y^2 ) \nonumber \\
&& + 2\pi i [123] (-\frac{2}{3} \pi^2 ) (G^3-\frac{3}{2}
 c_v  G^4 y ) + 2\pi i [1234] (\frac{i}{3}~\pi^3 )G^4 + 2\pi i [H]
 (\frac{\pi^2}{3} ) (G^3-\frac{3}{2} c_vG^4 y ) \nonumber \\
&& + 2\pi i ([H1] + [H2]) (-\frac{~i\pi^3}{3}G^4 ) + 2\pi i ([H3]-[H4])
 (-\frac{~\pi^2}{3}G^4 y^2 )
\eea
This clearly shows  that the perturbative series for each irreducible isospin
factor can be summed into a closed form  which is equivalent to
attaching to each vertex a modified coupling constant $\tilde{G} = G/(1 -
{1\over 2}c_v G\ln s)$.
Note that this modification of the coupling constant $G$,
which effectively modifies the propagator through Eq.(2), resembles
the Regge behavior of the gluon propagator in the conventional perturbative
QCD. In the abelian case where $c_v =0$ and $H$-type isospin factors
disappear, the amplitude can be summed into $ T-1 = 1-\exp(-2i\pi G) $ which
manifests the eikonalizing behavior.

To compare our result with that of
previous perturbative QCD calculations\cite{ChengWu}, we may rewrite
the result in the momentum space where the coupling $G$ becomes
\bea
 \int d^2 \rho G(\rho )e^{i\Delta \rho } &=&
{ g^2 \over 2\pi \Delta^2 } \nonumber \\
\int d^2 \rho G^{2}(\rho )e^{i\Delta \rho } &=&
\left({g^2 \over 2\pi }\right)^2 \int {d^2 q \over (2\pi )^2 }
{ 1\over q^2 (\Delta - q )^2 } \equiv {g^4 \over 4\pi^2 }I(\De) \\
\int d^2 \rho G^{3}(\rho )e^{i\Delta \rho } &=&
\left({g^2 \over 2\pi }\right)^3 \int {d^2 q_{1} \over (2\pi )^2 }
{d^2 q_{2} \over (2\pi )^2 }{ 1\over q_{1}^{2}q_{2}^{2}
(\Delta - q_{1}-q_{2} )^2 } \equiv {g^6 \over 8\pi^3 }I_{1}(\De) \nonumber \\
\int d^2 \rho G^{4}(\rho )e^{i\De\rho } &=& \left({g^2 \over 2\pi }\right)^4
\int {d^2 q_{1} \over (2\pi )^2 }
{d^2 q_{2} \over (2\pi )^2 }{d^2 q_{3} \over (2\pi )^2 }
{ 1\over q_{1}^{2}q_{2}^{2} q_{3}^{2}
(\Delta - q_{1}-q_{2} - q_{3})^2 }\equiv {g^8 \over 16\pi^3 }I_{2})(\De),
\nonumber
\eea
so that the amplitude is given by
\bea
\frac{i s}{2m^2}(T-1) &=&
-~\frac{sg^2}{2m^2}\frac{1}{\Delta^2} (1+
\overline{\alpha}\ln s+\overline{\alpha}^2\frac{I_1}{\Delta^2 I^2}\ln^2s
+\overline{\alpha}^3\frac{I_2}{\Delta^4 I^3}\ln^3s  )[1] \nonumber \\
& +& ~\frac{is}{4m^2} (g^4I+\frac{ng^6}{2\pi}\ln s~I_1+
\frac{3n^2}{16\pi^2}g^8\ln^2s~I_2 )[12] \\
&-& ~\frac{sg^6}{24m^2} (I_1+\frac{3n}{4\pi}g^2\ln s~I_2 )[H] +
{}~\frac{sg^6}{12m^2} (I_1+\frac{3n}{4\pi}g^2\ln s~I_2 )[123]
\nonumber \\
&-&~\frac{isg^8}{48m^2}I_2([1234] -[H1]-[H2])
-~\frac{sg^8}{48m^2\pi}\ln s~I_2 ([H3]-[H4]),\nonumber
\eea
where $\overline{\al}(\De)=\frac{3g^2}{4\pi}\De^2 I(\De)$
This agrees with previous perturbative calculation up to contributions
coming from
diagrams neglected in the QEU approximation.  It is remarkable that our
effective action within the QEU approximation shows eikonalizing and
Regge behaviors. We emphasize again that even though our
effective action has propagators of two dimensional nature, the action itself
can not be reduced to two dimensional effective action unlike previous
attempts\cite{Verlinde}. It is an outstanding open problem to include
interaction terms outside of the QEU approximation within the effective
action considered here.

\vskip 0.2in
{\bf Acknowledgements} \par
We would like to thank Kyungsik Kang, Chung-I Tan, and H. Fried
for discussions. We also thank Jin-Mo Chung for his valuable help with
the Feynman diagrams.
This work is supported in part by KOSEF(Q.P.), by CTP/SNU (S.N.),
by Non Directed Research Fund, Korea Reseach Foundation, 1993 (S.N.),
by BRSI-94-2442, and by the Kyung Hee Univ. Fund.


\end{document}